\documentclass[prl,twocolumn,showpacs,notitlepage,groupedaddress]{revtex4-1}
\usepackage{amsmath}
\usepackage{pdfsync}
\usepackage{graphicx}
\usepackage{braket}

\newcommand{\ionstate}[4]{\ket{\text{#1};\text{#2}_{{#3},{#4}}}}
\newcommand{\ion}[2]{$^{#1}${#2}$^+$} 
\newcommand{\half}{\frac{1}{2}}
\newcommand{\threehalf}{\frac{3}{2}}
\newcommand{\fivehalf}{\frac{5}{2}}

\begin{document}
	
\title{Atomic Combination Clocks}
\begin{abstract}
	 Atomic clocks use atomic transitions as frequency references. The susceptibility of the atomic transition to external fields limits clock stability and introduces systematic frequency shifts. Here, we propose to realize an atomic clock that utilizes an entangled superposition of states of multiple atomic species, where the reference frequency is a sum of the individual transition frequencies. The superposition is selected such that the susceptibilities of the respective transitions, in individual species, destructively interfere leading to improved stability and reduced systematic shifts. We present and analyze two examples of such combinations. The first uses the optical quadrupole transitions in a \ion{40}{Ca} - \ion{174}{Yb} two-ion crystal. The second is a superposition of optical quadrupole transitions in one \ion{88}{Sr} ion and three \ion{202}{Hg} ions. These combinations have reduced susceptibility to external magnetic fields and blackbody radiation.
\end{abstract}

\author{Nitzan~Akerman}
\author{Roee~Ozeri}
\affiliation{AMOS and Department of Physics of Complex Systems, Weizmann Institute of
	Science, Rehovot 7610001, Israel}
\maketitle

Atomic clocks are composed of three main components. An electromagnetic oscillator, an atomic reference to which the oscillator locks; i.e. a narrow atomic transition, and a frequency counter which translates the number of oscillator periods to a time unit. The atomic reference is usually carefully chosen to have as little as possible coupling to varying conditions in the environment which can lead to systematic frequency shifts. These systematic shifts can be generally divided into two categories. The first is due to the susceptibility of the atomic transition to different kinds of environmental electromagnetic fields such as the ambient magnetic field or blackbody radiation \cite{Rosenband2006Blackbody,Safronova2011Blackbody,dube2013evaluation}. The second kind of clock shifts are relativistic and emanate from the fact that the atoms are not investigated in an inertial frame. Here examples include second-order Doppler shifts or the gravitational red-shift \cite{chou2010optical}. Relativistic shifts can be only mitigated by maintaining the atoms in an inertial frame. The first kind however can be reduced not only by trying to minimize background electromagnetic fields, but also by reducing the differential susceptibility of the clock transition to different electromagnetic perturbations. The blackbody radiation shift, for example, can be reduced by choosing a particular atomic transition with a small differential polarizability  \cite{Rosenband2006Blackbody}. Alternatively, it was recently proposed to suppress blackbody radiation shift by using two different atomic clocks and a frequency comb to generate a ``synthetic'' frequency which is largely insensitive to the presence of electric fields \cite{yudin2011atomic}.

Reducing the susceptibility of atomic superpositions to electromagnetic perturbations has been the focus of intensive research in the context of quantum computing. Here, the motivation is to improve the coherence of multi-qubit superpositions for the purpose of metrology. One of the most successful methods for prolonging the coherence times of superpositions is the use of decoherence-free subspaces  \cite{lidar1998decoherence}. In this approach entangled superpositions of different states have reduced susceptibility to noise, leading to multi-second coherence times \cite{Roos2006Designer,Kotler2014Measurement}.  

We propose to use several atomic transitions in different species, each ticking with it own environmentally-sensitive frequency. These transitions are chosen such that their susceptibilities compensate each other in an entangled superposition leading to significantly reduced sensitivity to environmental fields. The idea of combining different materials with opposite susceptibilities in clocks dates back to the work of John Harrison in the eighteenth century \cite{usher2013history}. Harrison revolutionized the fields of timekeeping and navigation by building a marine chronometer based on a bi-metallic balance spring to compensate for temperature variations of the spring constant.

The phase of a multi-species entangled state will evolve at a frequency which is a linear sum of the different frequencies of individual transitions. It is thus possible to construct an atomic clock superposition that will evolve at a frequency of an ultra-violate transition, however, no ultra-violate laser will be involved in the clock operation. Instead, the superposition phase would be compared to the linear sum of frequencies of the clock lasers of the different transitions using an optical frequency comb. Fig.\ref{experimentScheme}(A) shows a schematics of our method. Inquiry the two-specie superposition phase with an optical frequency comb serves as a ''mechanical differential'' in analogy with the device used in vehicles which allows two wheels to rotate at different rotation frequencies while keeping the average rotation frequency fixed.  

% relating to multi-ion maximally  entangling state
Maximally entangled multi-atom states can, in principle, be used to surpass the frequency estimation accuracy attainable with uncorrelated atoms (standard quantum limit) and reach the Heisenberg limit \cite{Wineland1994Squeezed}. However, as pointed by Huelga et.al \cite{Huelga1997Improvement}, in practice, clock stability is ultimately limited by the finite coherence time of the system. For uncorrelated Markovian noise the two cases will give the same result. Nevertheless, in the some cases of correlated noise, one can sill gain by cleverly engineering the proper entangled state \cite{Roos2006Designer,Kessler2014heisenberg,Louchet2010entanglement}.  

%\section{Experimental implementation}
%________________________________________________________________________________________
\section{The Combination clock}

\begin{figure}[t!]
	%\begin{center}
		\includegraphics[width=8.5 cm]{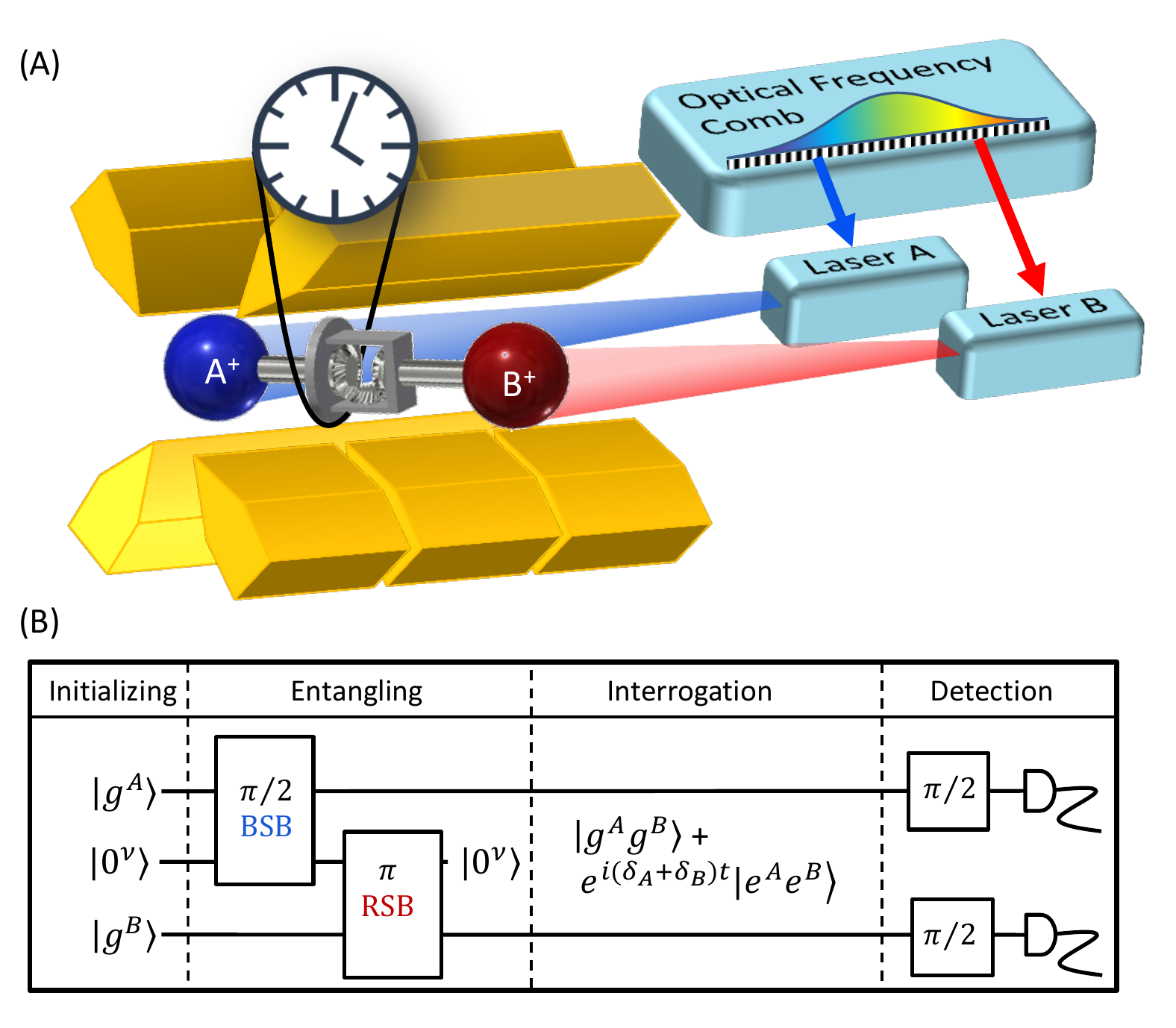}
		\caption{(A) schematic illustration of a two-species combination clock. The clock operates using the two clock lasers of the individual transitions. An optical frequency comb is used to compare the phase of the frequency sum of the two lasers to that of an entangled superposition. (B) The quantum circuit of gates for preparation of the clock entangled state and the final parity check allowing for clock interrogation. }
		\label{experimentScheme}
%	\end{center}
\end{figure}

We begin by describing our general scheme and discuss in detail the quantum protocol that can be used in the case of a two-ion combination clock . Our scheme is a generalization of the Ramsey spectroscopy method similar to the case of identical ions in a maximally entangled state \cite{Bollinger1996Optimal,Leibfried2004Toward}.

An $N$-ion,  multi-species, atomic register is prepared in a superposition of two states $\Psi=\frac{1}{\sqrt{2}}(\ket{G}+e^{i\Omega t}\ket{E})$. Where $\ket{G}=\prod_{i}^{N}\ket{f^{k,m}_i}$ and $\ket{E}=\prod_{i}^{N}\ket{s^{k,m'}_i}$ are tensor products of states in the ground and excited state manifolds of the different ions. Here $s$ and $f$ can be either the ground or excited states of the clock transition in ion $i$ of species $k$, and $(m,m')$ are the quantum numbers of the exact state in the ground or excited states manifolds of the transition (e.g. Zeeman or hyperfine state). The clock frequency is $\Omega=\sum_{{i}}^{N}\omega^{k,m,m'}_i$, where $\hbar\omega^{k,m,m'}_i = E(\ket{s^{k,m}_i}) - E(\ket{f^{k,m'}_i})$ and $E$ is the energy of the state. Notice that with our definition $\omega^{k,m}_i$ can be negative if ion $i$ is prepared such that $\ket{f^{k,m}_i}$ is the excited state and $\ket{s^{k,m}_i}$ is the ground state of the clock transition. The total frequency is thus a linear sum of the clock transition frequencies of all the ions in the crystal. 

After an interrogation time $t=\tau$ the phase between the atomic superposition and corresponding linear sum of local oscillators (i.e. clock lasers), $\delta\tau$, is obtained in the following way. A $\pi/2$ carrier pulse is applied to all ions in the crystal with their corresponding clock lasers followed by state detection and parity analysis. The probability of measuring even parity (i.e. an even number of ions in the excited state) will be given by $P_{even} = \cos{\delta\tau}$. A following change to the frequency of one of the clock lasers can lock the linear sum of frequencies to the entangled superposition.

Comparing with the case of a single clock-ion species, the use of multi ion-species increases experimental complexity due to the need of having a set of all the lasers, including narrow-linewidth clock lasers, that are necessary for each of the ion-species. On the other hand, some of the technical aspects such as single-ion addressing and detection are simplified owing to the spectral distinguishably between species. Comparing with logic spectroscopy schemes \cite{Schmidt2005Spectroscopy,Chou2010Frequency}, the combination clock does not involve motion-sensitive operations after the initial preparation stage and therefore continuous cooling or very-low heating rates are not a prerequisite for implementing this scheme.
% In the case of two different transition we have a rotating frame for each of ions. The acquired phases are with respect to the carrier transitions. 

The experimental sequence for the specific example of a two species two-ion crystal is depicted in Fig.\ref{experimentScheme}(B). The entanglement of two different ion species was recently demonstrated \cite{Tan2015multi,Ballance2015hybrid}. As a simple example, we begin with the two ions in the ground state of their internal state  $\ket{g^A},\ket{g^B}$ as well as in the ground state of one of their normal modes of motion $\ket{0}$. After applying a blue-sideband (BSB) $\pi/2$-pulse to ion A, we entangled the internal state of ion A with the motion of the two-ion crystal. Next we apply a red-sideband (RSB) $\pi$-pulse  on ion B which entangles the internal states of the ions while disentangling the internal states from motion. In the rotating-frame with respect to the sum of atomic transition frequencies $\omega_A+\omega_{B}$, the resulting state is,   

\begin{align}
\ket{g^Ag^B}&\ket{0} \overset{BSB}{\underset{\pi/2}{\Longrightarrow}}
					 \frac{1}{\sqrt{2}}\left(\ket{g^Ag^B}\ket{0}+\ket{e^Ag^B}\ket{1}\right)\\
					&\overset{RSB}{\underset{\pi}{\Longrightarrow}}\frac{1}{\sqrt{2}} \left( \ket{g^Ag^B}+e^{i\phi_{tot}}\ket{e^Ae^B} \right)\ket{0}\nonumber
\end{align}

Here $\phi_{tot}= \phi_{0}+(\delta_A+\delta_B)t_{int}$ with $\delta_{A,B}$ as the detunings of the two lasers from their corresponding atomic transitions and $\phi_{0}$ is a constant phase for zero interrogation time resulting from finite pulse times, light shifts, etc.  
The two-ion entangled state acquires a phase which is the sum of the two transition frequencies times the interrogation time $\tau$. This phase difference with respect to the sum of both clock laser frequencies is measured by a $\pi/2$ carrier pulse on each ion with its own clock laser, followed by state detection and parity analysis.

%\begin{align}
%Parity&=\left<\prod_{i}(\ket{S_i}\bra{S_i}-\ket{D_i}\bra{D_i})\right> \\
%	   &= \cos^2(\phi_{tot}/2) - \sin^2(\phi_{tot}/2)=\cos(\phi_{tot})\nonumber
%\end{align}

The combined two-ion clock is equivalent to a single atom with $\omega_0=\omega_A+\omega_B$ and with measured error signal $\delta=\delta_A+\delta_B$. In the case of two uncorrelated lasers it is enough to feedback on only one of the lasers while making sure that each of the lasers is sufficiently close to resonance such that it can execute the necessary pulses for the protocol; i.e. to within their Rabi frequency. One way to experimentally retrieve the stable clock frequency is by sum frequency generation using a nonlinear crystal. A more practical way is to work with correlated lasers which may be derived from a frequency comb. In this case we can write the two lasers frequencies as $\omega_A=N\cdot\omega_R$ and $\omega_B=M\cdot\omega_R$  where $N$ and $M$ are integers and $\omega_R$ is the comb repetition rate. Here the comb offset frequency is assumed to be set to zero. The error signal from the experimental interrogation $\delta_0=\delta_R(n+m)$ is then fed-back to stabilize $\omega_R$ to the value that locks $\delta= 0$. The extension of the locking of the comb to the linear sum in the case of more than two species is straightforward.
.     

%assuming that the free running laser has drifts which are much smaller than $1/\tau_p$ where $\tau_p$ is the pulse duration in the experimental sequence. Here the frequency stability is only manifested in the sum of the two lasers frequencies and not for each one separately.% 	THIS IS TRUE FOR ANY CLOCK INTEROGATION - EVEN FOR A SINGLE LASER. iN BETWEEN INTEROGATION CYCLES IT MUST BEHAVE NICELY
 
Below we analyze two examples where the combination of two optical transitions with differential scalar polarizabilities of opposite sign such that the BBR shift is considerably reduced. This is one of the more difficult systematic shifts to mitigate because, for a given polarizability, it requires the evaluation of the BBR fields with high accuracy. As shown below, other differential shifts due to electromagnetic moments can be either reduced (e.g. first order Zeeman) or enhanced (e.g. electric quadrupole shifts). It should be mentioned that this method is not limited to the combination of optical transitions only. It is also possible to combine optical clock transitions with a microwave hyperfine transition in an entangled superposition. Here the resulting effective clock frequency is very similar to that of the optical component, however the magnetic susceptibility of the combination superposition can be controlled, especially when working at a high magnetic field in the Paschen-Back regime. In this regime the effective g-factor of the hyperfine transition can be smoothly tuned to null the superposition first order magnetic susceptibility. 

We note that the use of entangled states is not always necessary. In many cases one can start with non-entangled states which have a significant overlap with the desired entangled state. If the part orthogonal to the desired clock superposition dephases quickly it will yield no parity signal. In this case, the measurement of parity will yield the same information as before, however at the cost of lower contrast and thus lower signal-to-noise ratio \cite{chwalla2007precision}. The advantage here would be the lower technical complexity as an entangling operation will not be necessary. 

%________________________________________________________________________________________
%\section{$^{40}Ca^+$ -  $^{174}Yb^+$ combination clock} 
\section{\ion{40}{C\MakeLowercase{a}} -  \ion{174}{Y\MakeLowercase{b}} combination clock} 
The first combination clock we examine is composed of the electric quadrupole transition $^4S_{1/2} \rightarrow$$ ^3D_{3/2}$ in \ion{40}{Ca} and the same electric quadrupole transition; $^6S_{1/2} \rightarrow$$ ^5D_{3/2}$; in \ion{174}{Yb}. This seems like a particularly nice combination since the two transitions are driven using fairly accessible wavelengths of $732\ nm$ and $436\ nm$ correspondingly; while the effective frequency at which this clock superposition rotates corresponds to an effective wavelength of 273 nm (1090\ THz) and lifetime  of $\approx 50$ ms limited by the \ion{}{Yb} transition. This effective wavelength is very close to that of the \ion{27}{Al} optical clock (267 nm) with the advantage that no UV light has to be used in order to drive this transition.  Furthermore the magnetic moments of these two transitions are the same to within a fractional difference of $10^{-3}$, so superpositions with states containing an equal number of states with opposite projections along the magnetic field direction can have almost zero magnetic susceptibility. But most importantly, the differential static polarizability of these two transitions are very similar in magnitude but have opposite signs \cite{Ludlow2015Optical} and therefore a superposition of the two species can have a significantly smaller susceptibility to blackbody radiation. Up to normalization, the specific superposition we consider here is,
\begin{equation}
	\ionstate{Ca}{S}{\half}{\half}\ionstate{Yb}{S}{\half}{-\half}+
	\ionstate{Ca}{D}{\threehalf}{\half}\ionstate{Yb}{D}{\threehalf}{-\half}.
\end{equation}
Below is an analysis of the different systematic shifts for this combination clock.

\subsection{First-order Zeeman}
The first-order Zeeman shift due to a uniform magnetic field for this superposition is almost zero as both superposition parts have a total zero magnetic moment, up to a small ($\approx10^{-3}$) difference between the bound electron $g$-factor in the two cases. Magnetic-field gradients, however, will still contribute a first order sensitivity to magnetic field. Here the magnetic field gradient can be measured, and subsequently nulled, by preparing the superposition, 
\begin{equation}
\ionstate{Ca}{S}{\half}{\half}\ionstate{Yb}{S}{\half}{-\half}+ \ionstate{Ca}{S}{\half}{\half}\ionstate{Yb}{S}{\half}{-\half}.				   
\end{equation}
A similar superposition has been previously used to estimate and compensate magnetic field gradients to the 20 mHz  level \cite{Kotler2014Measurement} which translates into a fractional uncertainty of $2\times 10^{-17}$. One can further improve on this uncertainty by interlacing experiments in which the spin direction of both ions is reversed, thus time-averaging out the gradient effect in the same manner that first order Zeeman shift is commonly averaged over \cite{Bernard1998laser}.
%-----------------------------------------------------------------------
\subsection{Second order Zeeman}
Other optical clocks such as  \ion{171}{Yb} benefit from the possibility of a clock transition which is first order insensitive to the magnetic field. In most of these cases this happens due to the hyperfine interaction between the electron and the nucleus. This first-order insensitivity then comes at the expense of a  relatively large second-order Zeeman shift (typically $10^3$ or more larger than in the first-order sensitive case). Here, since the first order insensitivity results from careful engineering of the clock superposition, the clock superposition amplitudes do not depend on the magnetic field value. A change in the magnetic field will not result in a change of the magnetic moment of the superposition states and hence no second order Zeeman shift. The small second order Zeeman shift will be, as in first order sensitive clocks, due to the fine-structure coupling in each of the transitions. The order of the second order Zeeman will be,
\begin{equation}
\Delta f_{M2} \approx \frac{(g\mu_B mB)^2}{h\Delta E_{FS}},
\label{secondZeemanEq}
\end{equation}
where $g$ is the electron dimensionless magnetic moment, $\mu_B$ is Bohr magneton, $B$ is the magnetic field, $h$ is Plank's constant and $\Delta E_{FS}$ is the fine-structure (FS) splitting in the D manifold. The FS splitting in \ion{}{Ca} is $\Delta E_{FS}/h = 1.82$ THz and in  \ion{}{Yb} it is  $\Delta E_{FS}/h =$ 42 THz. This means that the second order Zeeman will be dominated by \ion{}{Ca}. Working at a field of $10^{-4}$ T  we get, 
\begin{equation}
\Delta f_{M2} \approx \frac{(1.4\times 10^6)^2}{1.82\times 10^{12}} \approx 2\ Hz,
\end{equation}
One can measure the average magnetic field by preparing the superposition,
\begin{equation}
\ionstate{Ca}{S}{\half}{-\half}\ionstate{Yb}{S}{\half}{-\half}+
\ionstate{Ca}{S}{\half}{\half}\ionstate{Yb}{S}{\half}{\half}
\end{equation}
Here it is likely that the magnetic field will be measured with at least $10^{-3}$ accuracy, allowing for a clock frequency error as low as 4 mHz which translates into a fractional uncertainty below $2\times 10^{-18}$.

%-----------------------------------------------------------------------
\subsection{Electric Quadrupole shift}
The quadrupole shift (QS) of this clock superposition will be due to the quadrupole moment of the $D$ level of both ions since the $S$ orbital has no electric quadrupole moment.
The quadrupole shift of  $\left|D_\threehalf,m_{j}\right\rangle$ level is given by \cite{Itano2006Quadrupole},
\begin{equation}
\Delta f_{Q}=\frac{1}{4h}\frac{\partial{E_{z}}}{\partial{z}}\Theta\left(D,\frac{3}{2}\right)\frac{5-4m_{j}^{2}}{4}\left(3\mbox{cos}^{2}\left(\beta\right)-1\right).
\end{equation}
Here, $\Theta\left(D,\frac{3}{2}\right)$ is the $D_{\frac{3}{2}}$ level quadrupole moment and $\beta$ is the angle between the quantization axis set by the magnetic field and the quadrupole axis set by the trap geometry, where we assume full radial (cylindrical) symmetry. In the case of two ions in the trap, the equilibrium positions of the ions are $\pm z_0$, symmetrically positioned around the potential minimum along the trap axial direction. This results in an electric field gradient that is three times larger than the case of a single ion.
% add the following as remark 
%In the case of \ion{}{Yb} and \ion{}{Yb} the quadrupole moment of the $D_{\frac{3}{2}}$ level is $\Theta\left(D,\frac{3}{2}\right) = 2.08(11)\ ea_0^2$ \cite{Ludlow2015Optical} and $ 1.34(10)\ ea_0^2$ \cite{Itano2006Quadrupole} respectively. Reference\cite{Itano2006Quadrupole} does not give a confidence and therefore the uncertainty in the last value is estimated by the difference from the $Sr^+$ quadrupole moment reported in \cite{Itano2006Quadrupole} to the value measured in \cite{Shaniv2016Atomic}.
Using $\frac{\partial{E_{z}}}{\partial{z}}= 1000 V/mm^2$ and taking the case where $\beta=0$ as a worst case scenario, the QS for the $|D, m=\pm \frac{1}{2}\rangle$ will be 
\begin{equation}
\Delta f_{Q}=\frac{ea_0^2}{2h}\frac{1000}{10^{-6}}( 2.08 + 1.34) = 1156 \ Hz.
\end{equation}
This result is rather large due to the constructive addition of the entangled superposition and the position of the ions in the trap. 
In order to calibrate the QS one can place the ions in the superposition,
\begin{equation}
\ionstate{Ca}{D}{\threehalf}{\half}\ionstate{Yb}{D}{\threehalf}{-\half}+
\ionstate{Ca}{D}{\threehalf}{\threehalf}\ionstate{Yb}{D}{\threehalf}{-\threehalf}
\end{equation}
Which, in the absence of magnetic moment evolves solely due to the QS \cite{Roos2006Designer}. In this case the QS will be factor of two larger as compared with that of the clock superposition above. Here a value of $\beta$ can be chosen to minimize (null) the QS. Moreover the QS can be eliminated by averaging over different magnetic sub-levels \cite{dube2013evaluation}. For the case of the D$_{3/2}$ level, only the m$_j=1/2$ and m$_j=3/2$ states are required.

\subsection{Blackbody radiation shift}
The blackbody shift is due to the scalar differential polarizability of the clock transition; $\Delta \alpha _s$. The blackbody radiation creates a varying electric field at the ion position $E_{rms}$. The shift of the clock transition is given by,
\begin{equation}
\hbar\Delta\Omega = \frac{1}{2}E_{rms}^2\Delta \alpha _s(1 + \eta),
\end{equation}
Here $\Delta \alpha _s$ is the differential static polarizability between the two clock transition states. The dynamic correction factor $\eta$ takes into account the effect of BBR fields not being static but rather peaked at around $10\ \mu m$ (30 THz) at 300 K. This dynamic correction factor is typically small ($\approx 10^{-1}-10^{-2}$) and we will therefore neglect it in the following. The effective differential static polarizability of the superposition above (for the purpose of frequency shift calculation) will therefore be the sum of the two polarizabilities of the two transitions involved. 

The differential polarizability for the above $4S_{1/2} \rightarrow 3D_{3/2}$ transition in \ion{}{Ca} is $\Delta\alpha_s = -44(1)\ a_o^3$ [$-73(2)\ \frac{Jm^2}{V^2}$] \cite{Arora2007Blackbody-radiation,Safronova2011Blackbody}. This value is similar to the \ion{}{Ca} $4S_{1/2}\rightarrow3D_{5/2}$ clock transition which is investigated as an optical clock in several labs \cite{Chwalla2009Absolute,Wolf2011Direct,Matsubara2012Direct,Huang2016Frequency}. The corresponding differential polarizability on the same clock transition in \ion{}{Yb} is  $\Delta\alpha_s = 42(9)\ a_o^3$ [$69(14)\ \frac{Jm^2}{V^2}$] \cite{Schneider2005Sub-Hertz}. The differential polarizability of the \ion{}{Ca}-\ion{}{Yb} combination clock is $\Delta\alpha_s = -2 (9)\ a_o^3$ [$-4(14)\ \frac{Jm^2}{V^2}$]. This is a significant reduction in the expected BBR shift of at least a factor of $5$ if not more. The expected shift to the clock transition at $T = 300\ K$ is $20 (70)\ mHz$ corresponding to a fractional uncertainty of $6\times 10^{-17}$ dominated by the uncertainty in the differential polarizability. With better measurement of the this polarizability and a $10\ \%$ accuracy in estimating the BBR field power, the BBR uncertainty of this clock will be in the $10^{-18}$ range. 

\subsection{Second order Stark shift due to trap fields}
Trap fields will induce frequency changes due to the differential polarizability of the transition. Trap fields are mostly due to uncompensated micromotion or inherent micromotion that results from ion thermal motion in the pseudo-potential. Here, since the trap fields have a well-defined direction both the scalar and the tensor part of the differential polarizability will play a role.

To the extent that the trap rf fields are identical on both ions then the scalar polarizability of the transition is suppressed by the same factor that is mentioned in the BBR section. A difference in the rf amplitude the two ions experience is likely. Firstly, any stray field orthogonal to the trap axis will tilt the two ion crystal and displace the ions into different rf regions. This is because the radial pseudo-potential depends on mass and therefore the two ions experience different radial restoring forces. Secondly, any inhomogeneous axial rf fields will lead to a difference too. However, with the minimization of micromotion this residual shift should be at least as small as in the case of each of the two clocks when operated on a single ion.

The tensor polarizability of \ion{}{Ca} is $-17.43(23)\times 10^{-41}\ Jm^2/V^2$ \cite{Safronova2011Blackbody}. The tensor polarizability of $Yb^+$ is $-13.6(2.2)\times 10^{-41}\ Jm^2/V^2$ \cite{Ludlow2015Optical}. Therefore, the combined tensor polarizability of this clock transition is $-31.0(2.2)\times 10^{-41}\ Jm^2/V^2$. The amplitude of rf fields that the ions experience can be minimized to the $10\ V/m$ level. The order of magnitude of tensor shifts expected is therefore in the few $10's\ \mu Hz$ corresponding to fractional uncertainty in the $10^{-19}$ level. Furthermore, the tensor shift depend on the $m$ level and magnetic field direction, here the projection of the electric field, rather than the electric field gradient along the magnetic field direction is used. This means that the same average of measurements along different magnetic field directions or different $m$ levels that nulls the quadrupole shift will null the tensor polarizability contribution as well. 

%-------------------------------------------------------------------------------

\section{\ion{88}{S\MakeLowercase{r}} - 3$\times$\ion{202}{H\MakeLowercase{g}} combination clock} 
%-------------------------------------------------------------------------------
The second combination we look into here is motivated primarily by the desire to reduce BBR shifts as much as possible by reducing the combination differential scalar polarizability. The differential polarizability $\Delta\alpha_s=-47.938(71)\times 10^{-41}\ Jm^2/V^2$ \cite{Ludlow2015Optical} of the $S_{1/2}\rightarrow D_{5/2}$ electric quadrupole transition at $674\ nm$ in \ion{88}{Sr} is about 3 times larger and opposite in sign than the differential polarizability $\Delta\alpha_s=15\times 10^{-41}\ Jm^2/V^2$ \cite{Ludlow2015Optical} of the same transition in \ion{202}{Hg} at $282\ nm$. Thus, a superposition of a single \ion{88}{Sr} and three \ion{202}{Hg} ions would significantly reduce $\Delta\alpha_s$ to $\approx-3\times 10^{-41}\ Jm^2/V^2$ . 
\begin{align}
	&\ionstate{Sr}{S}{\half}{\half}\ionstate{Hg}{S}{\half}{-\half}\ionstate{Hg}{S}{\half}{-\half}\ionstate{Hg}{S}{\half}{\half}+\\
    &\ionstate{Sr}{D}{\fivehalf}{\half}\ionstate{Hg}{D}{\fivehalf}{-\half}\ionstate{Hg}{D}{\fivehalf}{-\half}\ionstate{Hg}{D}{\fivehalf} \nonumber
\end{align}
The effective frequency at which this superposition rotates is $82\ nm$ ($3634\ THz$) - deep in the extreme UV. The expected shift to the clock transition at T=300 K is below 20 mHz corresponding to a fractional uncertainty of $5\times10^{-18}$. With better measurement of the differential polarizability and 10\% accuracy in estimating the BBR field power, the BBR uncertainty of this clock will be in the $10^{-19}$ range. 

The rest of the systematic shift analysis here is rather similar to the one presented above for the \ion{40}{Ca} - \ion{174}{Yb} combination clock. In short:
The first-order Zeeman shift due to a uniform magnetic field for this superposition is zero as both superposition parts have a total zero magnetic moment. Moreover, with this particular spatial ordering of spin states, there would not be first order sensitivity to magnetic field gradients as well and only second order spatial magnetic field variation would cause a shift. The FS splitting in \ion{}{Hg} is $\Delta E_{FS}/h = 451$ THz and in \ion{}{Sr} it is $\Delta E_{FS}/h = 8.4$ THz and therefore second order Zeeman shifts will be dominated largely by  \ion{}{Sr} and can therefore be estimated using Eq. \ref{secondZeemanEq} to be $\Delta f_{M2} \approx0.2$ Hz. Here as well, an entangled superposition can be used to measure the average magnetic field on the clock array. Evaluating it at the $10^{-3}$ level, leads to a systematic shift of the order of $2\times 10^{-19}$. The quadrupole shift in this case will be rather large due to the large crystal and is estimate to be around 5 kHz for the same conditions considered above.

In conclusion, we propose and analyze the use of entangled states composed of multi ion-species for realizing an optical clock with reduced systematic shifts and  improved stability. As concrete examples we proposed two different combinations. A \ion{40}{C\MakeLowercase{a}} -  \ion{174}{Y\MakeLowercase{b}} two ions clock and a \ion{88}{S\MakeLowercase{r}} - 3$\times$\ion{202}{H\MakeLowercase{g}} four-ion crystal. Both these examples are expected to operate with a significantly reduced BBR shift. Our approach opens a new dimension to choosing suitable atomic clock references. Instead of merely scanning the periodic table for a suitable transition in different atomic species, we propose to search for the suitable linear combination of transitions, in different species which would have the lowest susceptibilities to environmental conditions. 

\bibliography{combinationBib}
\bibliographystyle{apsrev4-1}
%\begin{thebibliography}{12}
%	
%\bibitem{Breit} G. Breit, Nature, 122, 649 (1928)
%\bibitem{Kotler2014} S. Kotler, N. Akerman, N. Navon, Y. Glickman and R. Ozeri, Nature 510, 376 (2014)
%\bibitem{Itano2000} W. M. Itano, J. Res. Natl. Inst. Stand. Technol. 105, 829 (2000)
%\bibitem{RMP2015} A. D. Ludlow, M. M. Boyd, J. Ye, E. Peik and P. O. Schmidt Rev. Mod. Phys. 87, 637 (2015)
%\bibitem{Itano2006} W. M. Itano, Phys. Rev. A 73, 022510 (2006)
%\bibitem{Ravid2016} R. Shaniv, N. Akerman and R. Ozeri, Phys. Rev. Lett. 116, 140801 (2016)
%\bibitem{Roos2006} C. F. Roos, M. Chwalla, K. Kim, M. Riebe and R. Blatt, Nature 443, 316 (2016)
%\bibitem{Safranova2007} B. Arora, M. S. Safranova and C. W. Clark, Phys. Rev. A 76, 064501 (2007)
%\bibitem{Safranova2011} M. S. Safranova and U. I. Safranova, Phys. Rev. A 83, 012503 (2011)
%\bibitem{Blatt2009} M. Chwalla et. al., Phys. Rev. Lett. 102, 023002 (2005)
%\bibitem{Ubachs2011}A. L. Wolf et. al., Optics Lett. 36, 49 (2011)
%\bibitem{Hosokawa2012} K. Matsubara et. al., Optics Express 20, 22034 (2012)
%\bibitem{Gao2016} Y. Huang et. al., Phys. Rev. Lett. 116, 013001 (2016)
%\bibitem{Peik2005} T. Schneider, E. Peik and Chr. Tamm, Phys. Rev. Lett. 94, 230801 (2005)
%\bibitem{James1997} D. f. V. James, Appl. Phys. B 66, 181 (1998)
%\end{thebibliography}

\end{document}